# Structure, Magnetic Properties and Spin-glass Behavior in La$_{0.9}$Te$_{0.1}$MnO$_3$

Guotai. Tan, P. Duan, S. Y. Dai, Y. L. Zhou, H. B. Lu, Z. H. Chen[*]

Laboratory of Optical Physics, Institute of Physics, Chinese Academy of Sciences, Beijing 100080, People$^/$s Republic of China.

## Abstract

In this study we report the structure, magnetic and electrical transport properties of pervoskite oxide La$_{0.9}$Te$_{0.1}$MnO$_3$. This is a novel material with the space group R-3c, which shows the spin-glass-like feature at low temperature and has a good colossal magnetoresistance behavior. The magnetoresistance ratio $MR = [\rho(0) - \rho(H)]/\rho(0)$ is about 51% at 200 K in a field of 4 T. The XPS measurement suggests that Te ions are in the Te$^{4+}$ state, while Mn ions may be in the 2+ and 3+ valence state.

[*] Correspondence should be addressed to: Prof. Z. H. Chen, Laboratory of Optical Physics, Institute of Physics & Center for Condensed Matter Physics, Chinese Academy of Sciences, P.O. BOX 603, Beijing 100080, P.R. China.
Email address: zhchen@aphy.iphy.ac.cn    FAX:86-10-82649531

## I. INTRODUCTION

Doped manganites of the form $R_{1-x}A_xMnO_3$, where R presents rare earth and A is divalent cation, have attracted broad widely interesting due to its colossal magnetoresistance (CMR) effect, which is important for fundamental physical research and its potential in device design.[1] These compounds have a distorted perovskite structure and usually undergo ferromagnetic-paramagnetic phase transition at low temperature accompanied simultaneously with metal-insulator transition. When an external field is applied, their resistivities are significantly reduced at the temperatures near the Curie temperature (Tc). This phenomenon has been traditionally understood on the basis of the double-exchange (DE) model and Jahn-Teller effects.[2-5] When La ions in the $LaMnO_3$ were replaced by a divalent element such as Ba, Sr and Ca and et al, proportional amount of $Mn^{3+}$, with electronic configuration $t_{2g}^3 e_g^1$, is substituted with $Mn^{4+}$ ($t_{2g}^3$) and form a mixed-valence state of $Mn^{3+}$-$Mn^{4+}$. The hopping magnitude of $e_g$ electron between spin aligned $Mn^{3+}$ and $Mn^{4+}$ ions via $O^{-2}$ rests on the angle ($\theta_{ij}$) between the neighboring spins, and relates to the applied magnetic field. In a hole doped CMR system, the mixed-valence state of $Mn^{3+}$-$Mn^{4+}$ is a key component for understanding CMR effect and the transition from a ferromagnetic metal to a paramagnetic semiconductor. Note that $Mn^{3+}$ is a Jahn-Teller ion while $Mn^{4+}$ is not. Therefore, another mixed-valence state could also be created through replacing $Mn^{4+}$ using a $Mn^{2+}$, which is a non Jahn-Teller ion, to form $Mn^{3+}$-O-$Mn^{2+}$ double exchange system as shown in the electron doped samples such as $La_{1.8}Y_{0.5}Ca_{0.7}Mn_2O_7$,[6-7] $La_{0.7}Ce_{0.3}MnO_3$,[8] and $La_{1-x}Zr_xMnO_3$.[9] These studies indeed demonstrated that ferro-paramagnetic and metal-insulator transition arose from the double change between $Mn^{3+}$-O-$Mn^{2+}$. In this work, we investigate the structural, magnetic and electronic properties of a new compound with mixed-valence state of $Mn^{3+}$-$Mn^{2+}$, which is $La_{0.9}Te_{0.1}MnO_3$, replacing the La ions in the parent compound $LaMnO_3$ using tetravalent cation $Te^{4+}$. Experimental results suggest that $La_{0.9}Te_{0.1}MnO_3$ has the characteristic of CMR effect and spin-glass state.

## II. SAMPLE PREPARATION AND EXPERIMENTAL TECHNIQUES

Polycrystalline samples of $La_{0.9}Te_{0.1}MnO_3$ (LTMO) were prepared by conventional ceramic techniques. Stoichiometric quantities of $La_2O_3$, $TeO_2$, and $Mn_2O_3$ power were thoroughly mixed and subsequently pressed into disc shape. Then the disc sample was heated to 700 $^oC$ for 24 hours in the air. After this, the preheated sample was hand-ground with a mortar and pestle, granulated and repressed into disc shape. The latter were calcined for 12 hours in air at 900 $^oC$; this cycle was repeated a minimum of twice and the sample was sintered at about 930 $^oC$ for 24 h in the flowing oxygen gas before cooled down to room temperature in the off-powered furnace. X-ray diffraction (XRD) data at room temperature using Cu $K\alpha$ radiation were collected from 20 to 80 degrees using DMAX2400 diffractometer. Within the experimental errors, La:Te:Mn ratio of 0.9:0.1:1 were detected by chemical analysis and iodometric titration. The magnetization measurements were carried out with a Quantum Design superconducting quantum interference device (SQUID) magnetometer in the temperature range of 5-300 K. The electrical resistivity was measured using the standard four-probe method. The measurement of X-ray photoemission spectroscopy (XPS) was done by EscaLab 220-IXL electronic spectrometer. All spectra were recorded using Al- $K\alpha$ radiation and the scanning step is 0.05 eV. The Binding energy was calibrated with respect to the C-1s peak 284.6 eV at room temperature. The background due to the secondary electrons has been subtracted. Then all spectra have been fitted by Gauss function.

**III. EXPERIMENTAL RESULTS AND ANALYSIS**

The XRD spectrum at room temperature revealed that the structure of the samples had a rhombohedral lattice with the space group R-3c. The structure parameters were refined as a=0.5520 nm using the DBW9411 program and the Mn-O-Mn bond angle were also determined to be 165.37° using the powercell program.

Fig.1 shows the resistivity versus temperature curves in the temperature range of 5 to 300 K at 4 T, 2 T, and 0 T, respectively. All the curves display metal-semiconductor (MS) transition behaviors. We defined the temperature Tp where the resistivity is maximum. Below Tp, the variance ratio $d\rho/dT$ of curves is positive, which indicates the metallic behavior of the sample; oppositely, above Tp, $d\rho/dT < 0$, which means the sample has the feature of insulator. According to Fig.1, Tp were achieved at about 220, 217, and 212 K in the field of 4, 2, and 0 T, respectively. Compared with other analogous compounds with $Mn^{4+}$- $Mn^{2+}$ mixed-valence state, our Tp is much higher than that in the $La_{0.9}Zr_{0.1}MnO_3$,[9] but similar to that in the $La_{0.7}Ce_{0.3}MnO_3$. In addition, when external field is applied, the resistivity is suppressed significantly and Tp shifts to a higher value. This suggest that external magnetic field facilitates the hopping of $e_g$ between the neighbor Mn ions, which agrees with the DE model.

The magnetoresistance radio (MR), defined as $\Delta\rho/\rho(0) = [\rho(0) - \rho(H)]/\rho(0)$, where $\rho(0)$ and $\rho(H)$ are the resistivity of the zero field and an applied field H, respectively, is also plotted in the inset in Fig.1. Calculation shows the MR ratio for 2 T and 4 T applied field are 50.7% at 200 K and 33.15% at 195 K.

The temperature dependences of the ZFC (zero field cooled) and FC (field cooled) response were measured under the applied field of 0.01, 0.1, and 1 T as shown in Fig. 2. The ZFC data were collected by first cooling the sample from 300 to 5 K in zero applied field, then by turning on the field and warming. The FC data by cooling from 300 to 5 K in the applied field and then by warming from 5 K in the same field. As shown in Fig. 2, a ferromagnetic-paramagnetic phase transition is also observed. The transition temperatures Tc, determined as the temperature at which dM/dT is a minimum, are 230 K, 232 K, and 236 K at 0.01 T, 0.1 T, and 1 T, respectively. Contrasted with Tp, Tc variance with external field is similar to Tp, but closer to the room temperature.

It is noteworthy that a bifurcation occurs between the FC and ZFC curves at the low temperature region in a applied field of 0.1 T and 0.01 T, especially obvious for the latter one. This phenomenon is similar to the spin-glass behavior observed for other compounds, such as $Y_{1-x}Sr_xMnO_3$,[10] $La_{0.7}Pb_{0.3}(Mn,Fe)O_3$,[11] and $(La,Dy)_{0.7}Ca_{0.3}MnO_3$.[12] The temperature of the bifurcation is usually defined as the freezing temperature $T_f$, Based on Fig. 2, $T_f$ can be deduced to be about 200 K and 35 K for 0.01 T and 0.1 T, respectively. The $T_f$ decreases with increasing the external magnetic field, which is one of the typical characteristics of the spin-glass state in manganites. To further elucidate the nature of spin-glass state, we presented the hysteretic loops at 5 K with and without an applied field as shown in Fig. 3(a), which exhibits hysteretic behavior. This is also a characteristic of spin-glass-like state.[13-14] Besides, the relaxation between magnetization and time, which represents the irreversible magnetization process of the spin-glass system, is shown in Fig.

3(b). The magnetizations of all the curves rise sharply at first and then go to saturation with time under the application of the field. This is a typical relaxation phenomenon that further supports our presumption that spin-glass-like behavior exists in $La_{0.9}Te_{0.1}MnO_3$.

The XPS measurement was carried out in order to obtain more electronic structure information of the sample. The corresponding valence band (VB) and the core level spectra of Mn2p, Mn3s and Te3d are shown in Fig. 4. As shown in Fig. 4(a), the VB line has a similar shape to other hole or electron doping compounds such as $La_{0.9}Sr_{0.1}MnO_3$[15] and $La_{0.7}Ce_{0.3}MnO_3$,[16] which displays the characteristic double-peak structure located at 5.7 eV and 2.8 eV respectively, as illustrated by Gauss fitting curve. Saith et al[17] pointed out the former peak was primarily attributed to Mn3d-O2p bonding states, while the latter one was associated mainly with O2p nonbonding states. The entire VB region exhibits the feature of Mn3d and O2p orbital hybridized extensively. The enlarged spectrum near the Fermi level ($E_F$) is presented in the inset of Fig. 4(a), from which, two points are worth discussig. First, there exists a low–intensity shoulder at about 1eV below $E_F$, which suggests the $e_g$ electron in $Mn^{3+}$ ions provides a contribution to the VB.[18] Second, the existence of nearly negligible intensity at $E_F$ and its position relative to the shoulder mentioned above indicate that the Fermi level in $La_{0.9}Te_{0.1}MnO_3$ locates at the tail of the conductivity band and it has the characteristic of electron-dope.

The core-level spectrum of Mn3s and its corresponding fitting curve given by solid are shown in Fig. 4(b), which consists of two peaks located at 88.62 and 83.01 eV, respectively. It is well known that the 3s core level of Mn ion exhibit the exchange splitting which results from the exchange interaction between the Mn3s and Mn3d electrons, and the splitting magnitude depends on the valence state of the Mn ions.[19] Fig.

4(b) shows that the value of Mn3s splitting is about 5.61 eV, which is intermediate between those of LaMnO$_3$ with Mn$^{3+}$ and of MnO with Mn$^{2+}$ but closer to the former. This indicates that the existence of Mn$^{2+}$ may be possible in this material.

In Fig. 4(c), the Mn2p core-level XPS spectrum of the LTMO (the sign is T) is compared to those of the La$_{0.82}$Sr$_{0.18}$MnO$_3$ (the mark is S). The latter will be described in detail in another report.[20] Although both spectra are very similar to one another and display the spin-orbit split 2p$_{3/2}$ and 2p$_{1/2}$ peaks located around 642 eV and 654 eV, respectively, a small difference is stood out by the difference spectrum which is produced by subtracting spectrum S from spectrum T as shown at the bottom of Fig. 4(c). A positive and a negative peak are obtained in the range from 640 eV to 645 eV. The former one locates at 643.35 eV corresponding to Mn$^{4+}$-2p$_{3/2}$ and the latter at 639.86 eV which is close to the BE value 640.6 eV of Mn2p$_{3/2}$ in MnO.[21] Note that S has only formally Mn$^{3+}$ and Mn$^{4+}$ ions; therefore, Mn$^{2+}$ ions contribution in the difference spectrum is attributed to LTMO.

Fig. 4(d) illustrates the Te3d core-level spectrum, which exhibits 3d$_{5/2}$ and 3d$_{3/2}$ spin-orbit coupling double peaks standing at 586.38 and 575.93 eV, respectively. The latter is close to the BE of Te3d$_{3/2}$ in TeO$_2$ (576.1 eV),[22] thus the Te ions in this compound are probably in 4+ valence state. For Te-doped manganites, when a part of La$^{3+}$ ions are replaced by Te$^{4+}$ ions, it is possibly that electrons from La$^{3+}$ ions will be transferred to Mn ions due to the requirement of electronic neutrality. Consequently, these electrons will occupy the empty e$_g$ band and convert Mn$^{3+}$ ions into Mn$^{2+}$ ions. This further supports our assumption mentioned above that this material has a characteristic of Mn$^{2+}$-Mn$^{3+}$ mixed-valence state.

## IV. CONCLUSION

In summary, VI element Te, for first time, is doped in lanthanum manganite pevoskite and forms a novel material with colossal magnetoresistance: $La_{0.9}Te_{x0.1}MnO_3$. This compound with the maximum MR value of 50.7% displays a metal-semiconductor transition at around 230 K. Furthermore, its spin-glass-like behaviors at low temperature was observed that have no been reported in other doped-$LaMnO_3$ compounds with tetravalence-doped ions. Our recent study indicate that the similar spin-glass behaviors exist also in other composition (such as x=0.04, 0.15, and 0.2 etc.) in this material, further work is going on.

## ACKNOWLEDGEMENTS

This study was surpported by a grant from the State Key Program No. G1998061412 of China.

# Figure Caption

**Fig.1** R vs T curves of the sample in applied field H=0, 2, 4T.

Inset: MR versus Temperature at H=2, 4T.

**Fig2.** The Temperature dependence of the field-cooled (FC) and zero-field-cooled (ZFC) response of the sample measured in external field H=0.01, 0.1, 1T ;

**Fig.3.** (a) ZFC and FC hysteretic loops of the sample at 5K
(b) the magnetization relaxation curves of the sample

**Fig.4**.XPS measurement of the sample:

(a).VB and Gauss fitting curves; Inset is enlarge plot near $E_F$;

(b).Mn3s core-level;

(c).Mn2p spectra for $La_{0.9}Te_{0.1}MnO_3$ (T), $La_{0.8}Sr_{0.2}MnO_3$ (S) and their difference spectrum (T-S);

(d) Te3d.

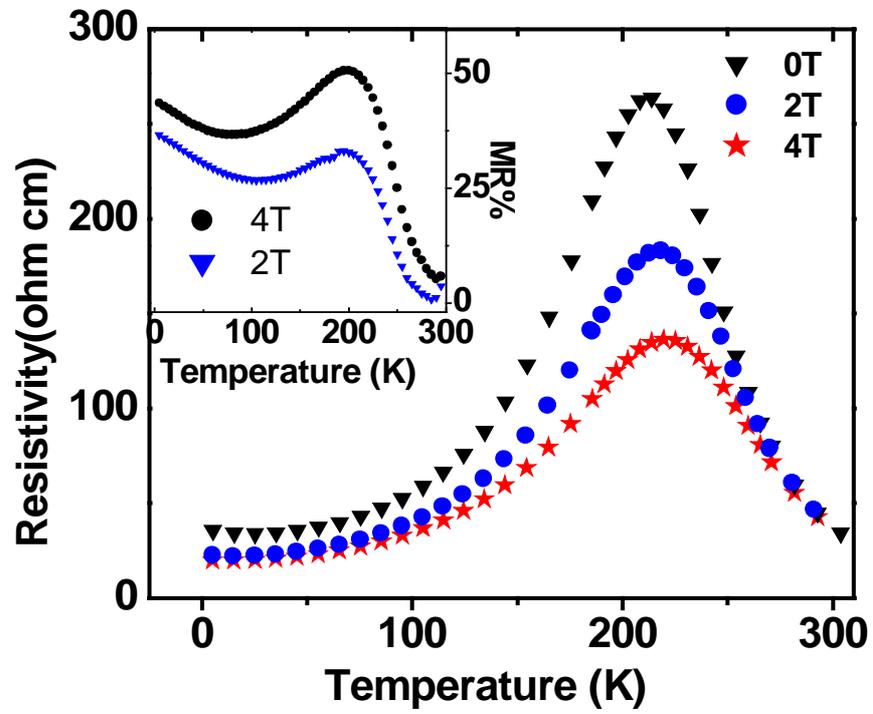

Fig.1. Tan et al

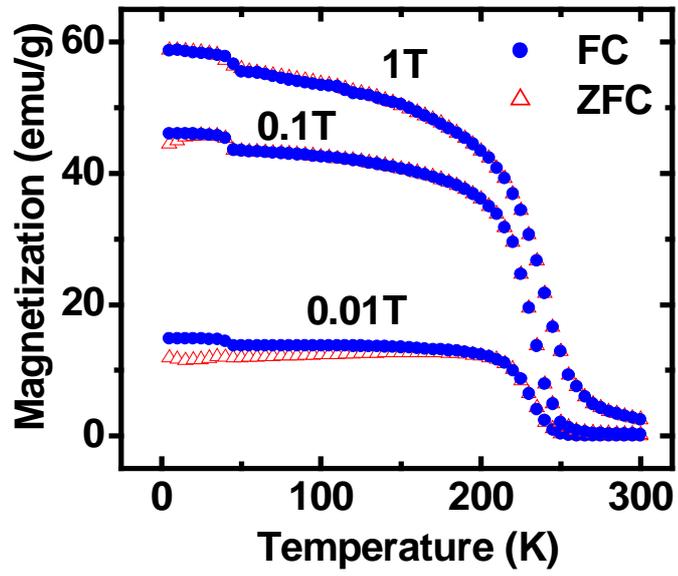

Fig.2. Tan et al

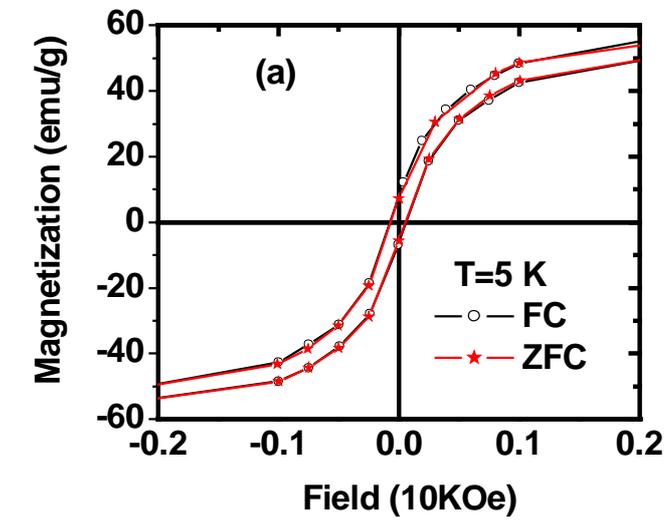

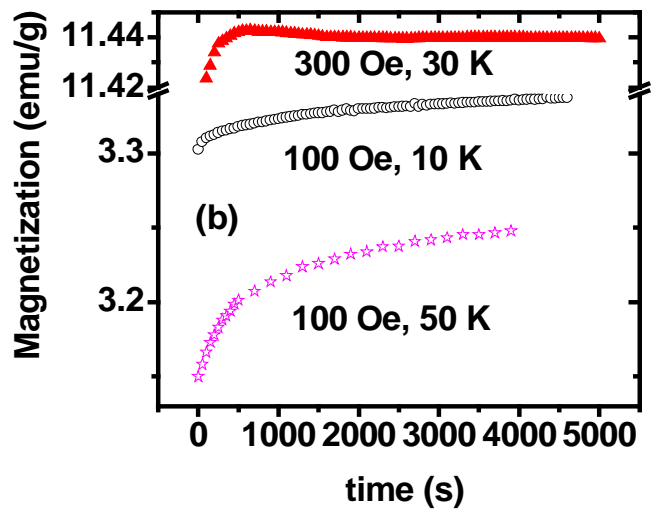

**Fig.3. Tan et al**

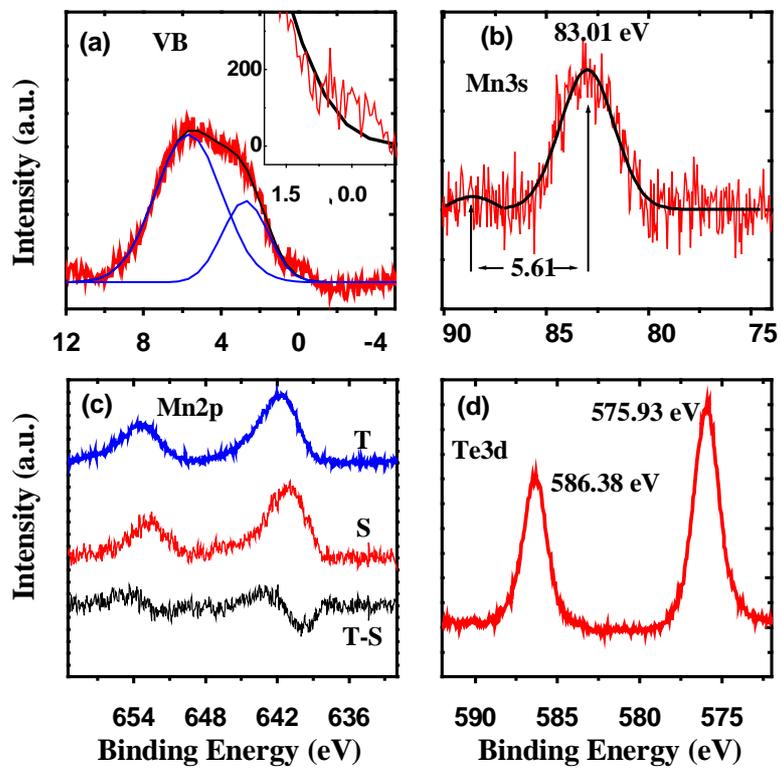

Fig.4. Tan et al